\documentclass[twocolumn,twocolappendix,resetfootnote]{aastex7}
\usepackage{CJK}
\usepackage{amsmath}
\usepackage{placeins}
\usepackage{graphicx}
\usepackage{subcaption}
\usepackage{hyperref}

\usepackage[utf8]{inputenc}
\usepackage[T1]{fontenc}
\usepackage{newunicodechar}
\usepackage{booktabs}  
\usepackage{threeparttable}
\usepackage{tabularx}
\usepackage{array}
\usepackage{soul, color, xcolor}
\newunicodechar{ʼ}{'}

\renewcommand\thefootnote{\arabic{footnote}}
\graphicspath{
  {images/}
  {images/corner/}
  {images/recovered signal/} 
  {images/4figures/}
  {images/poly fit/}
}

\begin{document}
\begin{CJK*}{UTF8}{gbsn} 
\title{Robust Extraction of Global 21 cm Spectrum from Experiments with a Chromatic Beam Based on Physics-Motivated Error Modeling}

\correspondingauthor{Yidong Xu}
\email{xuyd@nao.cas.cn}

\author[0009-0003-7621-1031]{Haoran Li (李浩然)}
\affiliation{State Key Laboratory of Radio Astronomy and Technology, National Astronomical Observatories, CAS, A20 Datun Road, Chaoyang District, Beijing, 100101, People's Republic of China}
\affiliation{School of Astronomy and Space Science, University of Chinese Academy of Sciences, Beijing 100049, People's Republic of China}
\email{lihr@bao.ac.cn}

\author[0000-0001-8075-0909]{Furen Deng (邓辅仁)} 
\affiliation{State Key Laboratory of Radio Astronomy and Technology, National Astronomical Observatories, CAS, A20 Datun Road, Chaoyang District, Beijing, 100101, People's Republic of China}
\affiliation{School of Astronomy and Space Science, University of Chinese Academy of Sciences, Beijing 100049, People's Republic of China}
\email{frdeng@bao.ac.cn}

\author[0000-0002-2744-0618]{Meng Zhou (周萌)}
\affiliation{State Key Laboratory of Radio Astronomy and Technology, National Astronomical Observatories, CAS, A20 Datun Road, Chaoyang District, Beijing, 100101, People's Republic of China}
\email{zhoumeng@bao.ac.cn}

\author[orcid=0000-0003-3224-4125,gname=Yidong,sname=Xu]{Yidong Xu (徐怡冬)}
\affiliation{State Key Laboratory of Radio Astronomy and Technology, National Astronomical Observatories, CAS, A20 Datun Road, Chaoyang District, Beijing, 100101, People's Republic of China}
\email{xuyd@nao.cas.cn}

\author[0000-0001-6475-8863]{Xuelei Chen (陈学雷)}
\affiliation{State Key Laboratory of Radio Astronomy and Technology, National Astronomical Observatories, CAS, A20 Datun Road, Chaoyang District, Beijing, 100101, People's Republic of China}
\affiliation{School of Astronomy and Space Science, University of Chinese Academy of Sciences, Beijing 100049, People's Republic of China}
\email{xuelei@cosmology.bao.ac.cn}



\begin{abstract}
The extraction of the sky-averaged 21 cm signal from Cosmic Dawn and the Epoch of Reionization 
faces significant challenges. The bright and anisotropic Galactic foreground, which is 4 – 5 orders of magnitude brighter than the 21 cm signal, when convolved with the inevitably chromatic beam, introduces additional spectral structures that can easily mimic the real 21 cm signal.  
In this paper, we investigate the signal extraction for a lunar-orbit experiment, where the antenna moves fast in orbit and data from multiple orbits have to be used. 
We propose a physics-motivated and correlated modeling of both the foreground and the measurement errors.
By dividing the sky into multiple regions according to the spectral index distribution and accounting for 
the full covariance of modeling errors,
we jointly fit both the foreground and the 21\,cm signal using simulated data for the Discovering the Sky at the Longest wavelength lunar orbit experiment. This method successfully extracts the 21 cm signals of various amplitudes from the simulated data
even for a testing antenna with a relatively high level of chromaticity. This approach, which is robust against moderate beam chromaticity, significantly relaxes the stringent design and manufacturing requirements for the antenna, offering a practical solution for future 21 cm global signal experiments either on the ground or in space.

\end{abstract}


\section{Introduction}

One of the most promising probes for the early cosmic history ($5 \lesssim z \lesssim 50$), including the Dark Ages, the Cosmic Dawn, and the Cosmic Reionization, is the H~{\small I} 21\,cm signal, which comes from the hyperfine transition of neutral hydrogen atoms.

The global 21 cm spectrum experiments seek to detected the sky-averaged redshifted 21 cm signal, which is in principle detectable with even a single antenna. Recent ground-based experiments include the Experiment to Detect the Global EoR Signature (EDGES \citealt{2008ApJ...676....1B}), the Shaped Antennas to measure the background RAdio Spectrum (SARAS \citealt{2013ExA....36..319P, 2018ApJ...858...54S}), the Probing Radio Intensity at high-Z from Marion (PRIZM, \citealt{2019JAI.....850004P}), the Sonda Cosmol\'{o}gica de las Islas para la Detecci\'{o}n de Hidr\'{o}geno Neutro (SCI-HI, \citealt{2014ApJ...782L...9V}), the Large aperture Experiment to detect the Dark Ages (LEDA, \citealt{2018MNRAS.478.4193P}) , the Broadband Instrument for Global HydrOgen ReioNisation Signal (BIGHORNS, \citealt{2015PASA...32....4S}), the Mapper of the IGM Spin Temperature (MIST, \citealt{Monsalve2024}), the Remote HI eNvironment Observer (RHINO, \citealt{Bull2024}), and the Radio Experiment for the Analysis of Cosmic Hydrogen (REACH, \citealt{2022NatAs...6..984D}). The EDGES claimed a positive detection of an absorption trough centered at 78 MHz with a depth of $\sim 0.5$ K \citep{Bowman2018}, which is much deeper than is expected by standard cosmological models \citep{Cohen2017MN, Cohen2020MN,Xu2021ApJ}. However, this result has not been confirmed by other experiments, and the SARAS3 has published a non-detection result \citep{SARAS32022}. The conflicting results between EDGES and SARAS3 indicate that there might be unknown systematics for ground-based experiments.

Some of the systematic effects in the ground-based experiments, such as ionosphere interference, ground reflection, and Radio Frequency Interference (RFI), can be avoided in space-borne experiments, especially if the experiment is performed on the lunar farside or lunar orbit. Some undergoing projects include the Hongmeng project (also known as the Discovering the Sky at the Longest wavelength or DSL, \citealt{2019arXiv190710853C, 2021RSPTA.37990566C}), the Dark Ages Polarimeter PathfindER (DAPPER, \citealt{2018JCAP...12..015T}), 
Probing ReionizATion of the Universe using Signal from Hydrogen (PRATUSH, \citealt{2023ExA....56..741S}), the Lunar Surface Electromagnetics Explorer (LuSEE-Night, \citealt{2023arXiv230110345B}), the Astronomical Lunar Observatory (ALO/DEX,\citealt{2024AAS...24326401K}), the Large-scale Array for Radio Astronomy on the Farside (LARAF, \citealt{2024RSPTA.38230094C}), and the CosmoCube \citep{Artuc2025}. 

The most challenging aspect of detecting the 21\,cm global spectrum is to extract the signal from the foreground, which is spectrally smooth but 4-6 orders of magnitude above the cosmological signal.
The structured 21 cm signal can be extracted by fitting the foreground with smooth functions such as polynomial models \citep{Bowman2018, Shi2022, Rao2017, Bevins2021}. Besides, one can also use singular value decomposition (SVD) of training sets to decompose foreground, signal, and other systematics into different eigenmodes and then extract signal (\citealt{Tauscher2018, Rapetti2020, Tauscher2020, Tauscher2021}). 

Inevitably, the beam pattern of any physically made antenna has some chromaticity, i.e. changes with frequency, which may add complex structures to the spectrum when convolved with the anisotropic intensity of the foreground, making the extraction of weak cosmological signal difficult \citep{Shi2022}. To mitigate the chromatic beam effect, \citet{Anstey2021, Anstey2023} proposed a physics-motivated modeling of the foreground and obtained robust fitting results. However, this method was only tested with ground-based experiments, where the sky changes slowly as the earth rotates. For a lunar orbit experiment onboard a fast-moving satellite such as the DSL mission, the result may be biased.

Here, we propose an improved modeling for measurement errors in addition to the foreground modeling, including the underlying covariance of measurements at different observation points in one or multiple lunar orbits. We adopt the sky division strategy in \citet{Anstey2021, Anstey2023} but include simulation-based covariance in the likelihood. Using mock spectra simulated with the expected parameters for the high-frequency daughter satellite of the DSL, which is dedicated to the global 21 cm spectrum measurement, we demonstrate the effectiveness of our method in avoiding bias in the extracted signal. 
We also discuss possible dependencies on configurations of observations and technical parameters.

This paper is organized as follows. In Section~\ref{sec:Simulation for the Global Spectrum Measurements on a Lunar Orbit}, we describe how we simulate the mock data of 21\,cm global spectrum, including the foreground sky, the 21\,cm signal model, and the beam models. In Section~\ref{sec:method}, we describe the uncorrelated foreground modeling used in \citet{Anstey2023} and the correlated modeling we develop in this work. In Section~\ref{sec:result}, we present the extracted 21\,cm global spectrum and discuss the dependences on foreground modeling, sky division, antenna beam, and signal amplitude. Finally, in Section \ref{sec:conclusion}, we draw our conclusions.

\section{Mock Data Simulation}
\label{sec:Simulation for the Global Spectrum Measurements on a Lunar Orbit}

In this section, we introduce how we simulate the mock spectra measured in a realistic lunar orbit, taking into account the anisotropic Galactic foreground, a simplified 21\,cm signal expected from the Cosmic Dawn (CD), beam models with different levels of chromaticity, time-varying Moon blockage, and the thermal noise for the experimental parameters as designed for the DSL mission.

\subsection{The foreground sky}\label{sec:foreground_sky}
\begin{figure}
    \centering
\includegraphics[width=0.9\columnwidth]{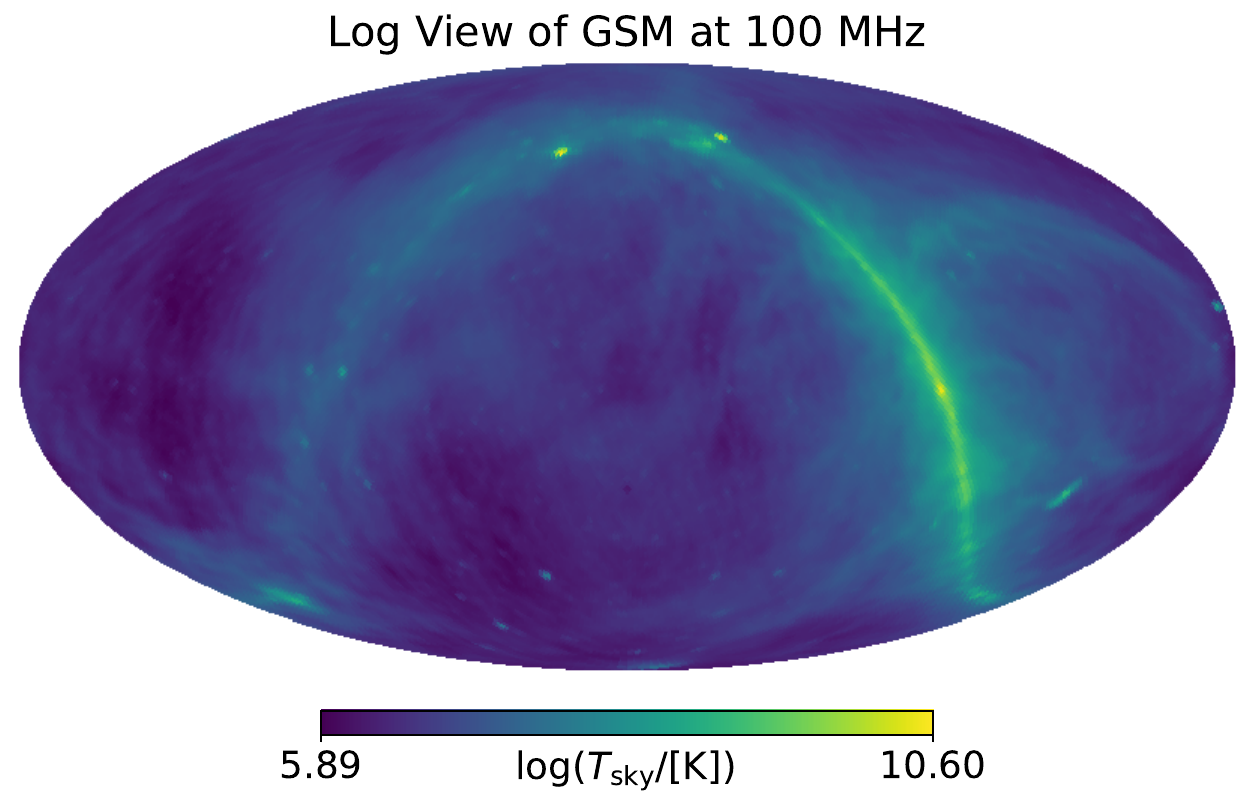}
\includegraphics[width=0.9\columnwidth]{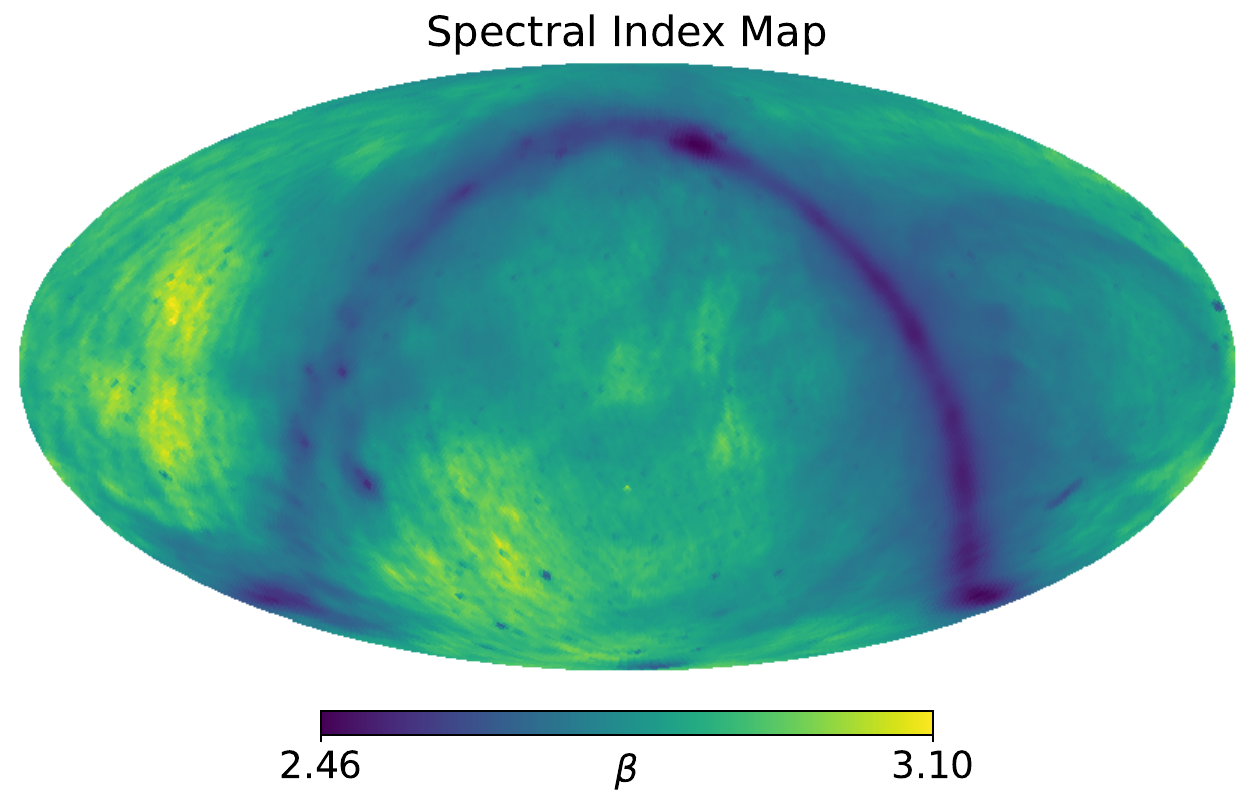}
    \caption{The brightness temperature map of the input foreground sky at 100 MHz
    (above) and the spectral index map derived from GSM408 and GSM230 (bottom) in the ecliptic coordinate system.}
\label{fig:brightness temperature and spectral index map}
\end{figure}

In order to ease comparison with previous methods (e.g., \citealt{Anstey2021,Anstey2023}), we use the same foreground model adopted in \citet{Anstey2021}, but apply it to the full frequency coverage
from 30 to 120 MHz with a bandwidth of 0.4 MHz.
The sky map is extrapolated from the Global Sky Model at 408 MHz (hereinafter GSM408) \citep{2008MNRAS.388..247D} using a power law specified for each pixel, and can be described as,
\begin{equation}
    T_{\rm{sky}}(\nu, \boldsymbol{\hat{n}}) = [T_{408}(\boldsymbol{\hat{n}})-T_{\rm{CMB}}](\frac{\nu}{408})^{-\beta(\boldsymbol{\hat{n}})}+T_{\rm{CMB}}\,.
    \label{eq:powerlaw skymap}
\end{equation}
Here $T_{408}(\boldsymbol{\hat{n}})$ denotes the GSM408 map (resolution $\sim$ 1 degree, $N_{\rm side}=64$ in the HEALPix pixelation scheme), $T_{\rm{CMB}}=2.726$ K is the effective temperature of the cosmic microwave background (CMB), and $\beta(\boldsymbol{\hat{n}})$ is the spectral index for each pixel, derived by comparing GSM408 with the 230 MHz sky map from the same model (hereinafter GSM230).
The sky map at 100 MHz and the spectral index map are shown in Figure \ref{fig:brightness temperature and spectral index map}. 

This sky model takes into account both the anisotropic brightness temperature distribution and the anisotropic spectral index distribution, which is appropriate for the investigation of the chromatic beam effect on the 21 cm signal extraction. In this work, we neglect the free-free absorption effect, which significantly affects spectral shape below $\sim$ 10 MHz \citep{ULSA2021}. 
In this paper, our focus is on signal extraction methods, with further investigations into the effects of alternative input sky models reserved for future work.

\subsection{The 21 cm signal}

We use a Gaussian profile for the input 21\,cm signal in our simulation, which can be described as,
\begin{equation}
    T_{\rm{21cm}}(\nu) = -A \exp[{-\frac{(\nu-\nu_0)^2}{2\sigma_0^2}}]\,,
\label{eq:signal}
\end{equation}
where $A$ is the amplitude, $\nu_0$ is the central frequency, and $\sigma_0$ is the characteristic width of the absorption trough.
In our fiducial simulation, we adopt $A=0.2$ K, which is the maximum signal level expected in standard cosmology \citep{Xu2021ApJ}, $\nu_0=75$ MHz, and $\sigma_0=6$ MHz.
We will also discuss the cases with different signal levels, i.e. 
$A=0.15$ and $0.1$ K, in Section \ref{sec:results_correlated_modeling}.

\subsection{Beam models}
We consider three kinds of antennas, all of which have rotationally symmetric beam patterns:
\begin{itemize}
\item[(a)] {\bf Dipole-like antenna.} 
It is a hypothetic monochromatic antenna similar to an ideal dipole antenna, but the beam pattern is the same as the ice-cream antenna (as described below) at 30 MHz.

\item[(b)] {\bf Ice-cream antenna.} 
It consists of a cone covered by a half-sphere with a total height of about 1.4 m, and an oblate cylinder with a diameter of about 1.6 m (Figure 3 in \citealt{VZOP2024DSL}). 
This antenna follows the current design for the high-frequency daughter satellite of the DSL and is being specially designed as an integrated satellite while being less chromatic.

\item[(c)] {\bf Disc-cone antenna (badly designed)\footnote{A well designed disc-cone antenna can have very low chromaticity, the example used in this case is a badly designed one with a too large disc, we use it for illustration. }.} It consists of a cone with a diameter of 40 cm and a height of 30.5 cm, and a disc reflector with a diameter of 200 cm. This is a test antenna with a relatively high level of chromaticity. 
\end{itemize}

\begin{figure*}[ht!]
    \centering
    \begin{subfigure}{0.4\textwidth}
        \centering
        \includegraphics[width=0.9\linewidth]{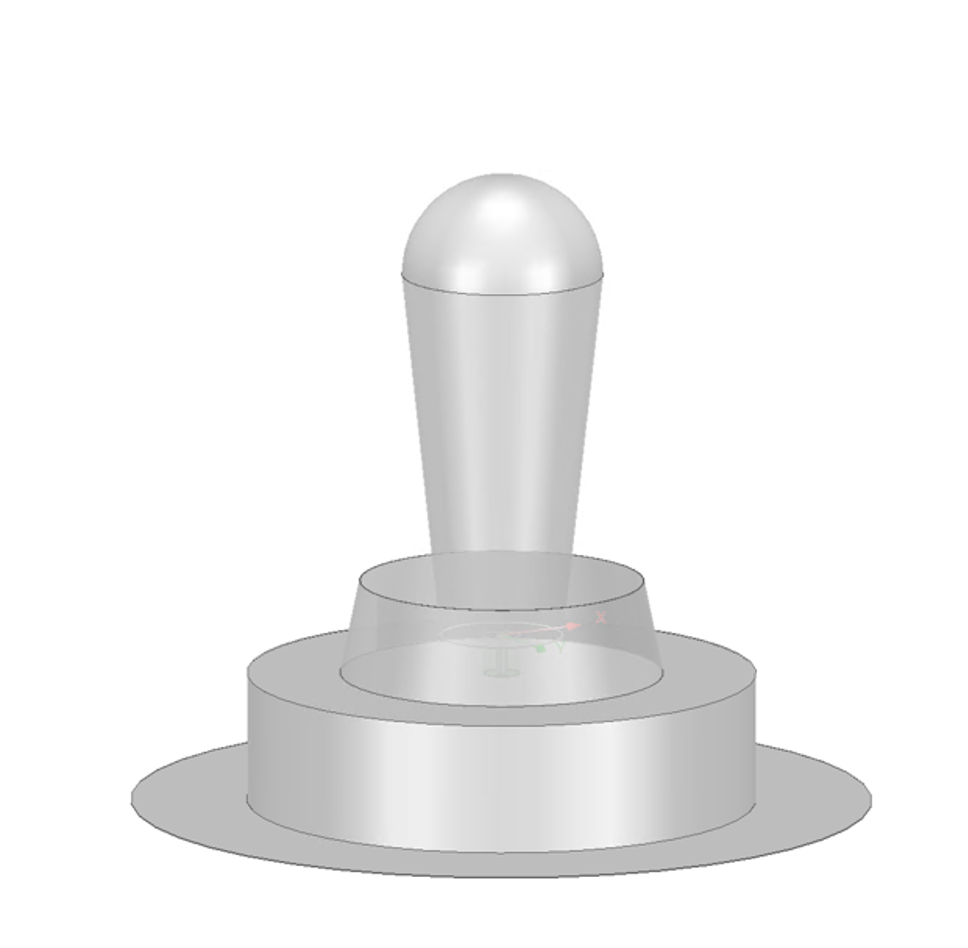}
        \caption{Ice-cream antenna}
    \end{subfigure}
    \hspace{0.01\textwidth}
    \begin{subfigure}{0.5\textwidth}
        \centering
        \includegraphics[width=1.0\linewidth]{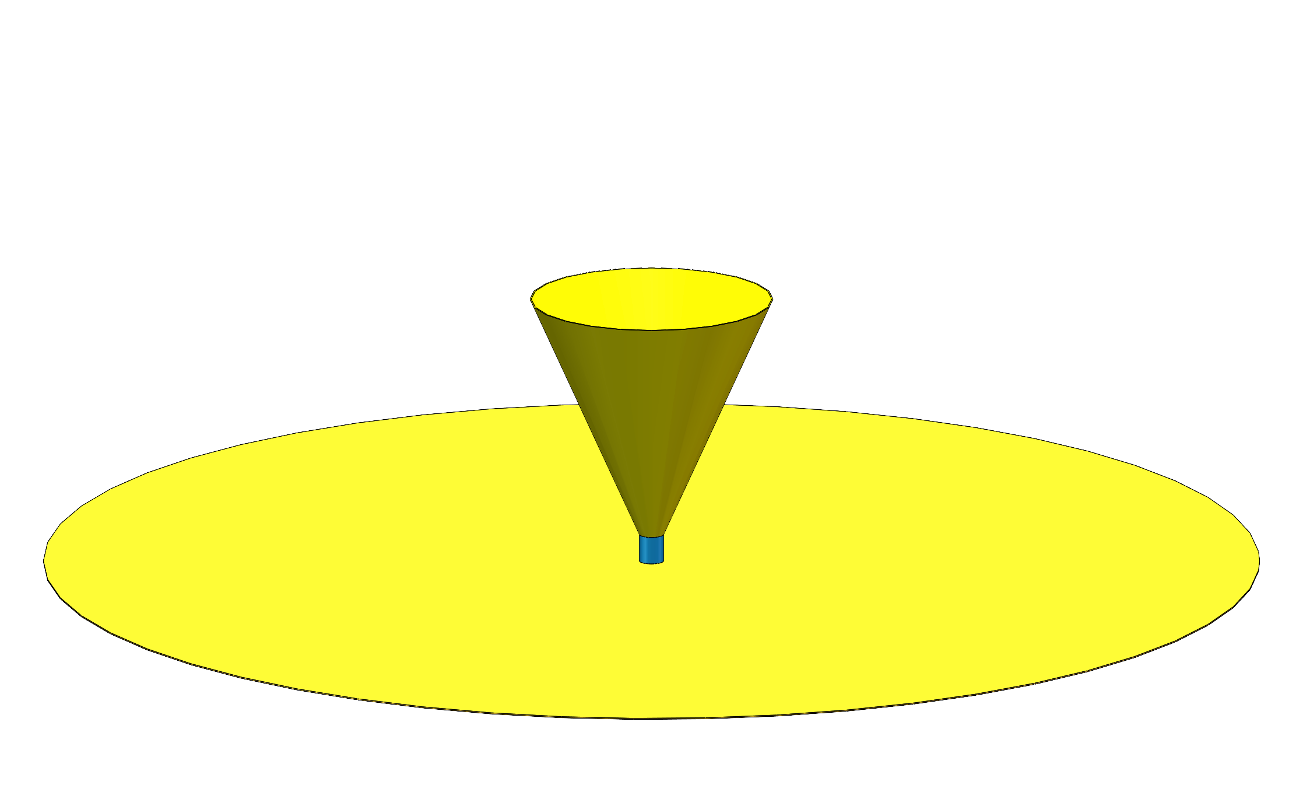}
        \caption{Disc-cone antenna}
    \end{subfigure}
    \caption{Diagrams of the ice-cream antenna (left panel) and the disc-cone antenna (right panel). The exact dimensions are subject to change during the detailed design phase.
    }
    \label{fig:diagram of two antenna}
\end{figure*}

\begin{figure*}[ht!]
    \centering
    \includegraphics[width=0.9\textwidth]{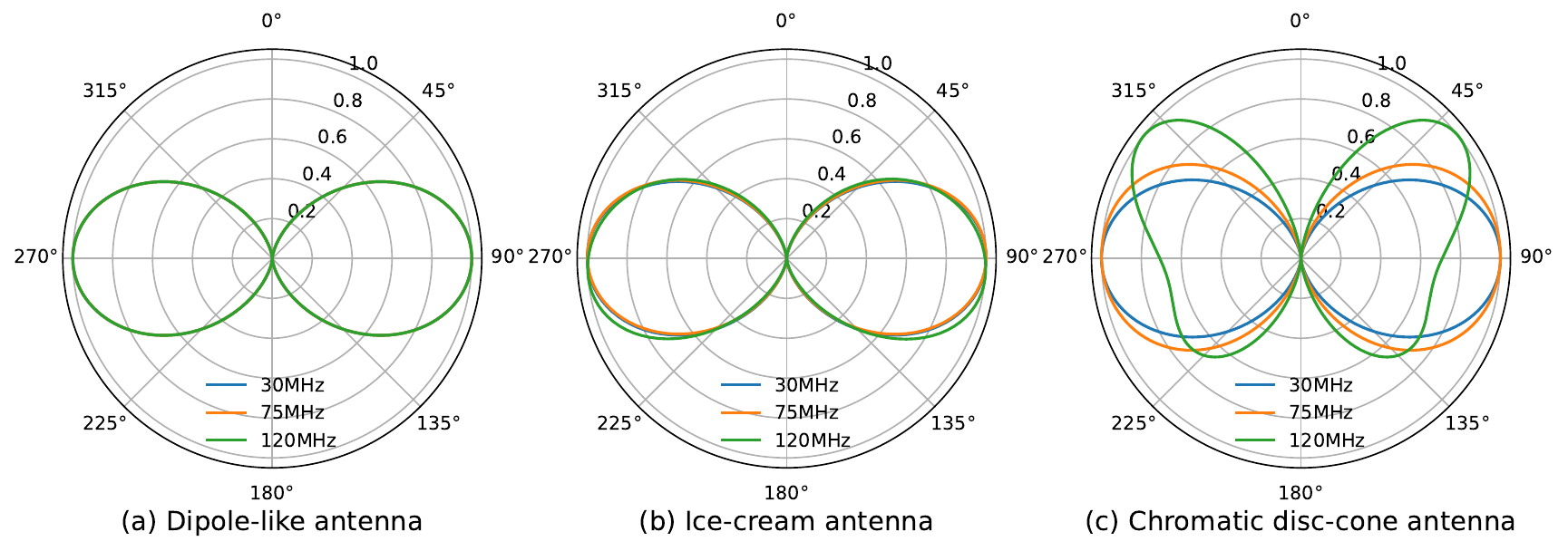}
    \caption{The cross sections of the simulated beam profiles of the dipole-like antenna (left panel), the ice-cream antenna (middle panel), and the chromatic disc-cone antenna (right panel), respectively, 
    at three different frequencies as indicated in the legend. 
    }
    \label{fig:beam comparison}
\end{figure*}

Diagrams of the ice-cream antenna and the disc-cone antenna are shown in the left and right panels of Figure~\ref{fig:diagram of two antenna}, respectively.
Figure \ref{fig:beam comparison} shows the cross sections of the simulated beam patterns at three different frequencies. Note that the rotational symmetry of the beam response is set for simplicity in both antenna design and simulation analysis. In practice, the real instrument may not achieve the perfect symmetry. However, slight deviation from the rotational symmetry will not impact our results as long as we can measure the beam precisely, which is technically feasible on the lunar orbit.

\subsection{Mock data}
Given the foreground model, the 21\,cm signal, and the beam model, the measured mock spectrum can be simulated by,
\begin{align}
    &T_{\rm{mock}}(t,\nu) = 
    \frac{1}{\int B(t, \nu, \boldsymbol{\hat{n}})S(t,\boldsymbol{\hat{n}})d\Omega}\times\bigg\{ T_{\rm{th}}(t,\nu)+\nonumber \\
    &\left. \int B(t, \nu, \boldsymbol{\hat{n}})S(t,\boldsymbol{\hat{n}})\,
    [T_{\rm{sky}}(\nu, \boldsymbol{\hat{n}})+T_{\rm 21cm}(\nu)]d\Omega  \right\},
    \label{eq:mock spectrum}
\end{align}
where $B(t, \nu, \boldsymbol{\hat{n}})$ is the antenna beam response, $S(t,\boldsymbol{\hat{n}})$ is the shade function that describes the Moon blockage, and $T_{\rm{th}}(t,\nu)$ is the thermal noise. 
In the DSL mission, the antenna dedicated to the global 21 cm spectrum measurement is kept aligned so that the beam null always points to the center of the Moon. So both the antenna response and Moon blockage should vary with time as the satellites orbit the Moon.
We assume $T_{\rm{th}}(t,\nu)$ is white noise with a variance $\sigma_{\rm{th}}(t, \nu)$, as given in \citet{Shi2022}. Note that here we consider the time dependence of thermal noise, because the system temperature varies with the sky temperature, as the Moon blockage varies with time.
The beam is normalized so that $\int B(t, \nu, \boldsymbol{\hat{n}})d\Omega = 1$.

In the DSL mission, the satellites move fast in a circular lunar orbit at a height of 300\,km with an inclination of $30^\circ$, the orbit period is 8248 seconds. To ensure that the sky seen by the antenna does not vary much within an integration time, we divide the orbit equally into 30 observation points. The time taken for the satellite moving across one segment is 1/30 of the period, i.e. 274.93 s. 
An integration time of 2500\,s on each of the 30 observation points can be achieved after about 9 orbits. With this assumed integration time on each observation position in orbit and a bandwidth of 0.4 MHz, the expected thermal noise level after integrating 30 measurements on 30 observation points is about 20 mK around 60 - 90 MHz, which would be subdominant to systematics in the signal extraction process. If we further take into account that we only use observations when satellites are shielded from the Earth ($\sim$ 1/3 of each orbit), we would need about 30 orbits.

\begin{figure}
    \centering
    \includegraphics[width=0.9\columnwidth]{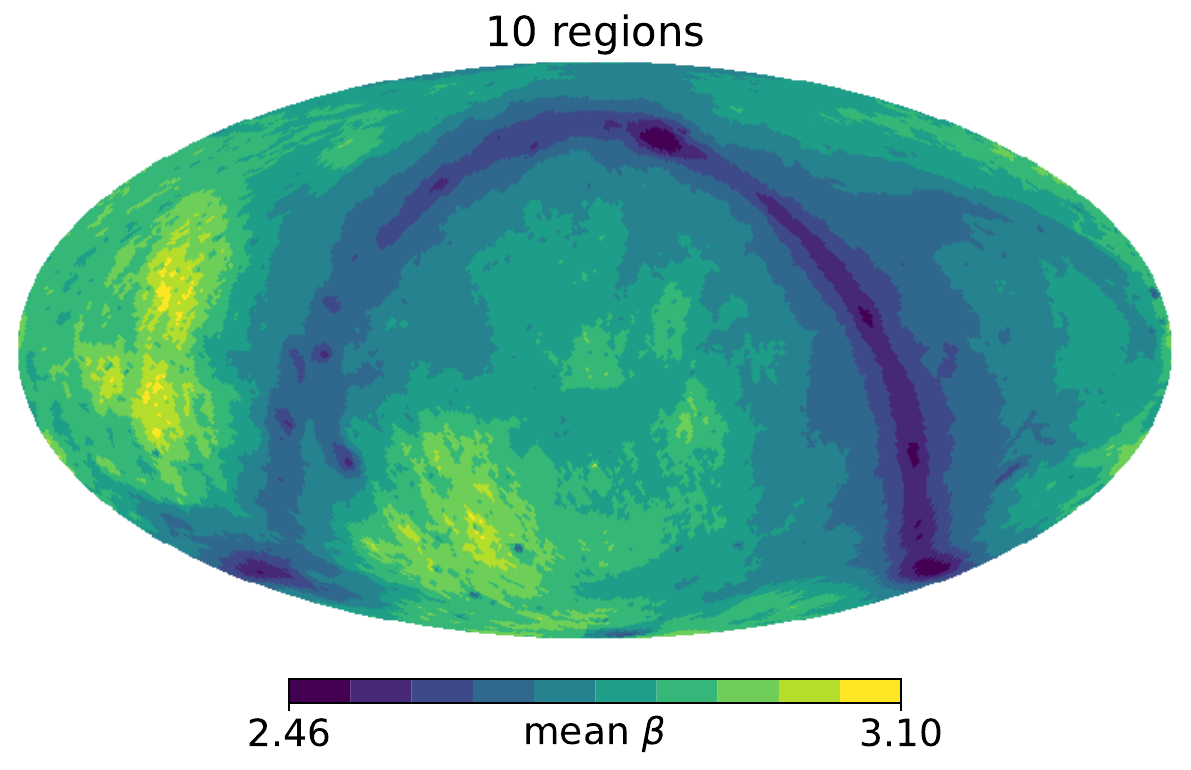}
    \includegraphics[width=0.9\columnwidth]{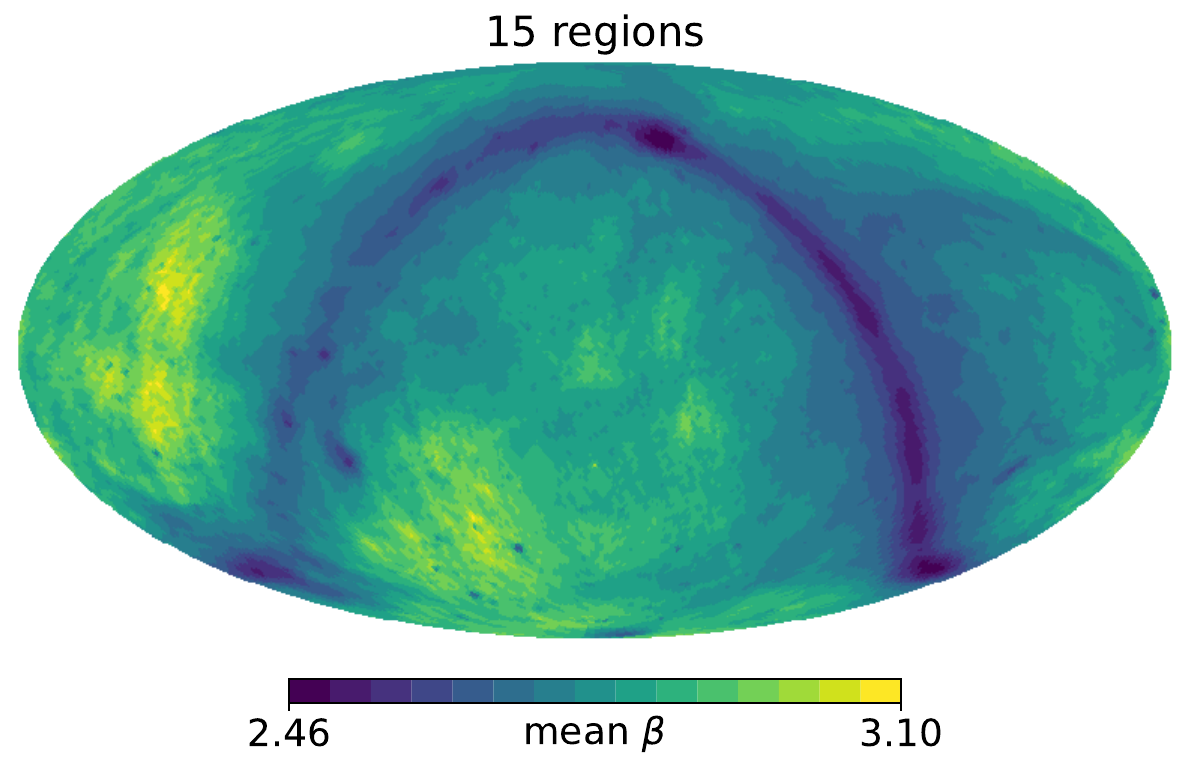}
    \includegraphics[width=0.9\columnwidth]{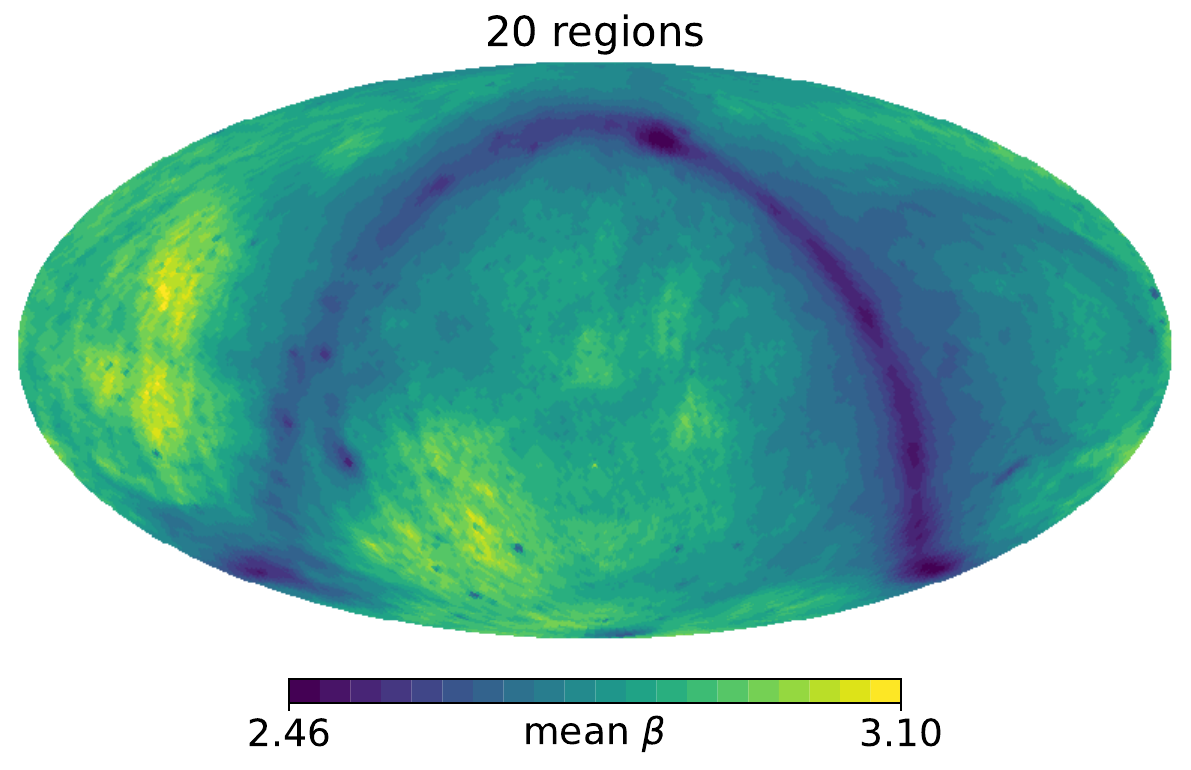}
    \caption{The spectral index map after the sky division. We divide the whole sky into 10 (upper), 15 (middle) and 20 regions (bottom), respectively.}
    \label{fig:divided sky}
\end{figure}

\section{Methodology}
\label{sec:method}
In general, one may hope to extract the 21\,cm signals from foregrounds by fitting the measured spectra with some parameterized functions. 
Assuming a beam pattern with weak chromaticity, \citet{Shi2022} used a five-term logarithmic polynomial as the foreground fitting model and successfully extracted the global 21\,cm signal. However, at such low frequencies, the beam chromaticity of antennas is not negligible. The inhomogeneity of the foreground sky, coupled with the chromatic beam, can result in additional structures on the measured spectrum and make the extraction biased.

In this work, to mitigate the effect of chromatic distortions, we extend the method proposed in \citet{Anstey2021,Anstey2023}, and develop a new physics-motivated modeling for both the foreground itself and the statistical errors in the modeled data.
We first review the original method in Section~\ref{sec:Signal Extraction using Physics-Motivated Foreground Modeling}  and then describe how we modify it with simulation-based covariance in Section~\ref{sec:improved_likelihood}.

\subsection{Uncorrelated Modeling}
\label{sec:Signal Extraction using Physics-Motivated Foreground Modeling}

As the spectral index of the foreground varies gradually, we divide the whole sky into $N_r$ regions, as an approximation, the variation of the spectral index within each region is neglected.  Illustrative cases of $N_r= $ 10, 15, and 20 are shown in Figure \ref{fig:divided sky}. The fitting model is thus expressed as,
\begin{eqnarray}
        T_{\rm{model}}(t,\nu)&=&\sum_p^{N_r}K_{p}(t,\nu)(\frac{\nu}{408})^{-\beta_p}\nonumber\\ &+& T_{\rm{CMB}} + T_{\rm 21cm}(\nu)\,,
    \label{eq:model spectrum}
\end{eqnarray}
where 
\begin{equation}
    \begin{aligned}
        & K_{p}(t,\nu) = 
        \frac{1}{\int B(t, \nu, \boldsymbol{\hat{n}})S(t,\boldsymbol{\hat{n}})d\Omega} \\
    &\times \int B(t, \nu, \boldsymbol{\hat{n}})S(t,\boldsymbol{\hat{n}}) M_p(\boldsymbol{\hat{n}})[T_{408}(\boldsymbol{\hat{n}})-T_{\rm{CMB}}]d\Omega\,.
    \end{aligned}
    \label{eq:K_N}
\end{equation}
Here $M_p(\boldsymbol{\hat{n}})$ is a mask function, with 1 for pixels in the $p$th region and 0 elsewhere, and $\beta_p$ is the constant spectral index in that region.
The corresponding Gaussian likelihood between the fitting model and the mock data is given by
\begin{equation}
    \begin{aligned}
        \ln\mathcal{L}=-\frac{1}{2}\sum_i^{N_t}\sum_j^{N_{\nu}}\left[\frac{T_{\mathrm{model}}(t_i,\nu_j)-T_{\rm{mock}}(t_i,\nu_j)}{\sigma^{\rm eff}_{\rm{th}}(t_i,\nu_j)}\right]^2\,,
    \end{aligned}
    \label{eq:initial likelihood}
\end{equation}
where 
\begin{equation}
\sigma^{\rm eff}_{\rm{th}}(t_i,\nu_j)=\frac{\sigma_{\mathrm{th}}(t_i,\nu_j)}{\int B(t, \nu, \boldsymbol{\hat{n}})S(t,\boldsymbol{\hat{n}})d\boldsymbol{\hat{n}}}\,.
\end{equation}
Here, $N_t$ and $N_\nu$ are the number of time-dependent observation points and the number of frequency channels, respectively. We neglect the constant term in the logarithm of the likelihood. For convenience, we define $\mathbf{y} = (\mathbf{T}_{\rm model}-\mathbf{T}_{\rm mock})/\sigma_{\rm th}^{\rm eff}$, which is the normalized residual between the fitting model and the mock data, and rewrite Equation~(\ref{eq:initial likelihood}) in the matrix form,
\begin{equation}
    \ln\mathcal{L}= -\frac{\mathbf{y}^T\mathbf{y}}{2}.
\end{equation}
In this model, $\mathbf{y}$ is assumed to be fully uncorrelated. However, $\mathbf{y}$ is expected to be correlated due to the time-varying response function (i.e., the product of $B$ and $S$) and the non-uniformity of the true sky map within each region \citep{Anstey2023,Hibbard2023ApJ}. Therefore, in the case of the DSL mission, we need a likelihood with correlations between residuals.

\subsection{Correlated Modeling}
\label{sec:improved_likelihood}
In this section, we introduce how we construct a likelihood with correlations between residuals. We assume no correlation between the thermal noise and model bias. The full covariance of $y(t,\nu)$ can be separated into the model bias contribution $\mathbf{C}_{\rm model}$ and the thermal noise contribution $\mathbf{C}_{\rm th}$. Therefore, the likelihood with correlations can be given by
\begin{equation}
    \begin{aligned}
        \ln\mathcal{L} &= -\frac{\mathbf{y}^T(\mathbf{C}_{\rm model}+\mathbf{C}_{\rm th})^{-1}\mathbf{y}}{2}
    \end{aligned}
    \label{eq:improved likelihood}
\end{equation}
$\mathbf{C}_{\rm th}$ is the identity matrix $\mathbf{I}$ since $y(t,\nu)$ is already normalized with the thermal noise level, $\mathbf{C}_{\rm model}$ is estimated numerically below.

The dimension of $\mathbf{y}(t,\nu)$ is $N_t\times N_\nu$, so $\mathbf{C}_{\rm model}$ is a huge matrix. To reduce the computation costs, we perform eigenvalue decomposition on $\mathbf{C}_{\rm model}$ and remove the components whose eigenvalues are lower than $10^{-11}$. The number of remaining components is $\sim$ 100.

\begin{figure}
    \centering
    \includegraphics[width=\linewidth]{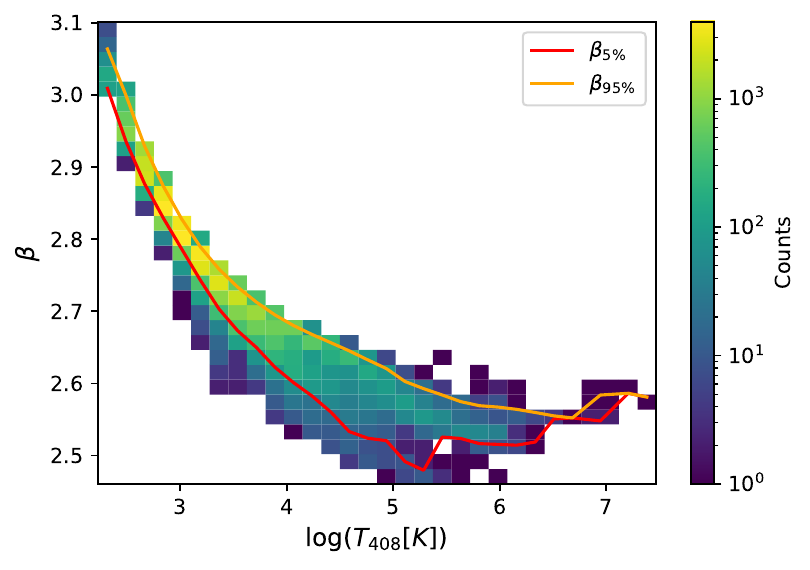}
    \caption{Two-dimensional histogram between logarithmic brightness temperature of GSM408 and spectral indices $\beta$. The red and orange lines represent $5\%$ and $95\%$ quantiles of spectral indices in each logarithmic bin, respectively.}
    \label{fig:2D_hist_and_lines}
\end{figure}

Now we estimate the covariance $\mathbf{C}_{\rm model}$ in a simulation-based way. 
To compute $\mathbf{C}_{\rm model}$, we have to statistically realize the mock data for a number of times, which should be more than the number of effective eigenvalues. In each of the numerical realizations, we take into account the correlation between the spectral index and the brightness temperature of the pixels according to the GSM408 map. We first group the brightness temperatures of pixels on the GSM408 map into 30 logarithmic bins with a bin size of 0.175 $\ln$[K]. The distribution of the spectral indices of pixels in each of the brightness temperatures bin is shown in the two-dimensional histogram in Figure~\ref{fig:2D_hist_and_lines}.
In each brightness temperature bin, we compute the 5$\%$ quantile $\beta_{5\%}$ and 95$\%$ quantile $\beta_{95\%}$ of the spectral indices, which are shown as the red and orange lines in the Figure \ref{fig:2D_hist_and_lines}, respectively. Then we interpolate them on each pixel based on its brightness temperature at 408 MHz.
In each of the realizations of mock data, we randomly sample the spectral index for each pixel assuming a uniform distribution between $\beta_{5\%}$ and $\beta_{95\%}$ quantiles. 
Although the real distribution of spectral indices is not uniform, this is a reasonable approximation that we can make without introducing any unknown physics or any extra parameters. 
Finally, we get mock spectrum measurements for 1000 times. Note that although the number of realizations is smaller than the dimension of $C_{\rm model}$, there are only around 100 effective eigenvalues ($>10^{-11}$) in the matrix, which means 1000 realizations should be enough.

We use the least squares method to fit these 1000 mock spectra based on the model of Equation~(\ref{eq:model spectrum}) and estimate the covariance by,
\begin{equation}
    \mathbf{C}_{\rm model}=\frac{\mathbf{R}^T \times \mathbf{R}}{1000}\,,
    \label{eq:covaiance calclation}
\end{equation}
where $\mathbf{R}$ denotes the residuals of 1000 fitting results normalized by the thermal noise. 

\begin{figure}
    \centering
    \includegraphics[width=0.9\columnwidth]{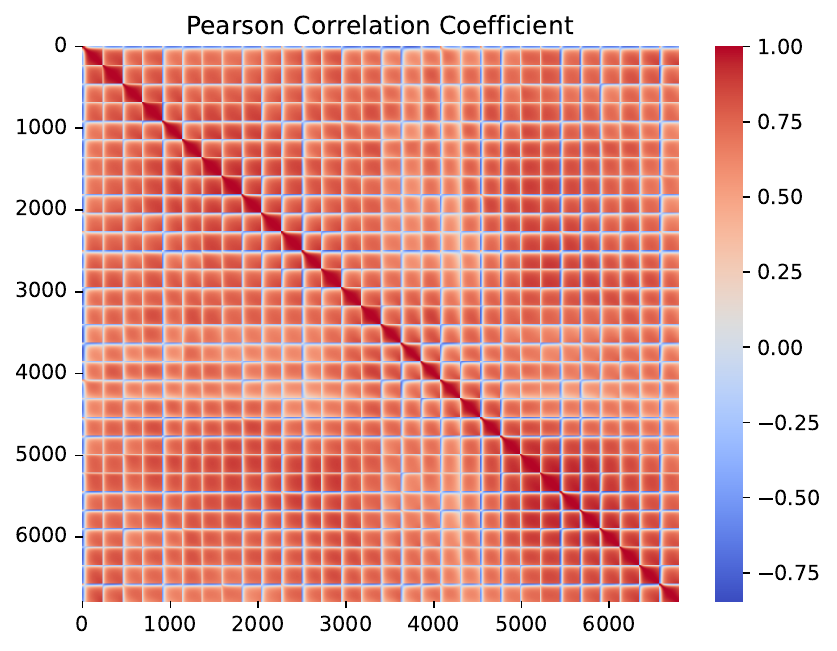}
    \caption{The normalized covariance (Pearson Correlation Coefficient) matrix of residuals for an observation with a disc-cone antenna and $N_r=20$ in one entire orbit. Apparently, the residuals are correlated.}
    \label{fig:Pearson Correlation Coefficient.png}
\end{figure}

In Figure~\ref{fig:Pearson Correlation Coefficient.png}, we plot the normalized covariance matrix for observations with the disc-cone antenna and $N_r=20$ in one entire orbit for illustrative purposes. 
The covariance characterizes the correlation of residuals between each frequency and each observation point. In order to calculate $\mathbf{C}_{\rm model}$, the residuals are flattened in observation point and frequency. With 30 observation positions in one orbit and 226 frequency channels from 30 -- 120 MHz, $\mathbf{R}$ has a dimension of $N_t\times N_\nu = 30 \times 226 = 6780$. In Figure~\ref{fig:Pearson Correlation Coefficient.png}, each square grid (with a dimension of $226\times226$) represents the correlation between residuals of a spectrum measured at one position and a spectrum measured at another position.
The non-zero values of off-diagonal elements of the covariance indicate that the residuals are correlated.
Using our model for the spectrum, the residuals are mostly negative at the 30 MHz end, and become positive at higher frequencies. This results in the negative correlation between the residuals of one spectrum at the 30 MHz end and another spectrum at higher frequencies. The sign change in residuals from one spectrum to another results in the visual gridlines appearing in the covariance matrix.

Putting aside the negative residuals at the low-frequency end around 30 MHz, the residuals are mostly positive, which result in the positive correlations between residuals of spectra measured at various positions. The small fluctuations in the correlation coefficients matrix are due to the anisotropic temperature distribution of the Milky Way.

\section{Results}
\label{sec:result}
Next, we show the extracted 21\,cm global spectra with both methods described above. We also explore how the goodness of extraction depends on the fitting modeling, sky division, antenna beam, and 21\,cm signal magnitude. 

We model the 21 cm signal as in Eq.~(\ref{eq:signal}), and set uniform priors for $A$ in the range of [0 K, 0.3 K], $\nu_0$ with [50 MHz, 100 MHz], $\sigma_0$ with [2 MHz, 10 MHz], and $\beta_p$ with [2, 4]. We use {\tt{POLYCHORD}}\footnote{\url{https://github.com/PolyChord/PolyChordLite}} \citep{2015MNRAS.450L..61H, 2015MNRAS.453.4384H} to sample these parameters. {\tt{POLYCHORD}} is a parallelized nested sampling algorithm designed for high-dimensional parameter spaces. It uses slice sampling at each iteration to sample within the hard likelihood constraint of nested sampling.

\begin{figure*}[htbp]
    \centering
    \includegraphics[width=0.96\textwidth]{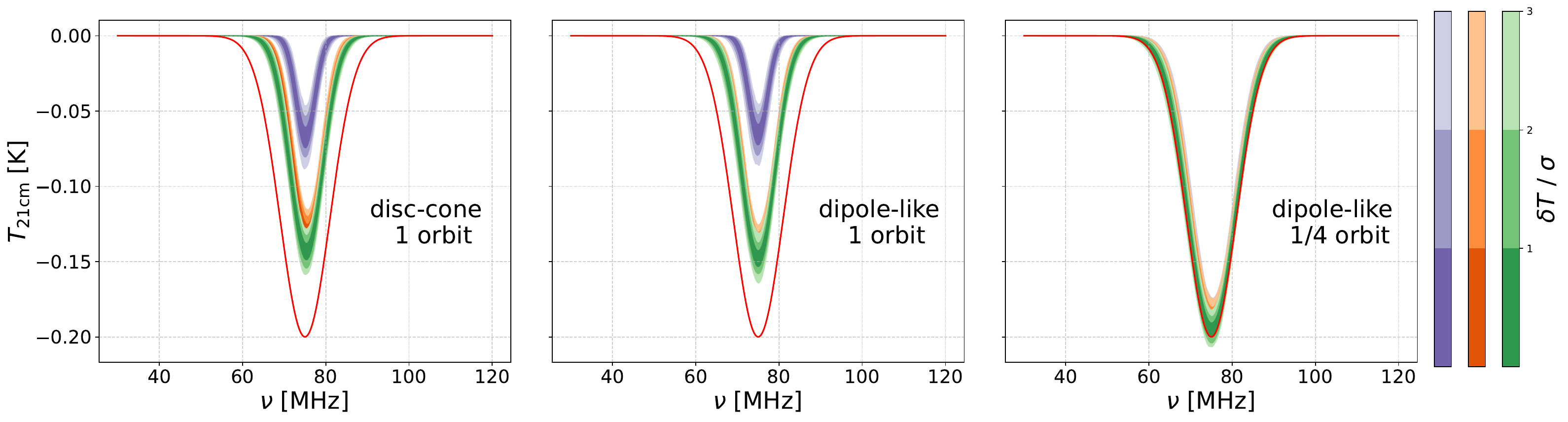}
    \caption{Extraction of the 21\,cm global spectrum using uncorrelated modeling. The left panel and the middle panel show
    the fitting results from mock data in one entire orbit using a disc-cone antenna and a dipole-like antenna, respectively. We also plot the result fitted from mock data in one quarter orbit convolved with a dipole-like antenna in the right panel. In each panel, the green, orange and purple shaded regions represent the recovered signals using models with $N_r=20$, 15 and 12 sky regions, respectively.  The red curve is the input 21\,cm signal, and the colored regions with different darkness show 1, 2, and 3$\sigma$ uncertainties of the recovered signals.}
    \label{fig:recovered_signal_nocov_comparison}
\end{figure*}

\begin{table*}[]
    \centering
    \caption{ Extractions of the 21\,cm global spectrum. The eight columns are the orbit, the antenna beam model, the input signal amplitude, the number of divided sky regions, the fitting modeling, the fitted signal amplitude, the fitted central frequency, and the fitted frequency width, from left to right respectively. $\nu_0$ and $\sigma_0$ of the input 21 cm signal are 75 MHz and 6 MHz, respectively.}
    
    \hspace{-6em}\begin{threeparttable}
    \begin{tabular}{c|c|c|c|c|c|c|c}
    \hline
        \hline
          Orbit & Antenna & $A_{\rm input}$ [K] & $N_{\rm r}$ & Model & $A$ [K]& $\nu_0$ [MHz]& $\sigma_0$ [MHz]\\
        \hline
        Entire & Disc-cone & 0.2 & 12 & Uncorrelated & $0.068^{+0.007}_{-0.007}$ & $75.123^{+0.273}_{-0.280}$ & $2.297^{+0.188}_{-0.169}$ \\
        Entire & Disc-cone & 0.2 & 15 & Uncorrelated & $0.131^{+0.005}_{-0.005}$ & $75.567^{+0.185}_{-0.185}$ & $3.727^{+0.151}_{-0.150}$ \\
        Entire & Disc-cone & 0.2 & 20 & Uncorrelated & $0.143^{+0.005}_{-0.005}$ & $75.336^{+0.180}_{-0.178}$ & $4.008^{+0.150}_{-0.147}$ \\
        Entire & Dipole-like & 0.2 & 12 & Uncorrelated & $0.066^{+0.007}_{-0.007}$ & $74.982^{+0.285}_{-0.284}$ & $2.265^{+0.192}_{-0.165}$ \\
        Entire & Dipole-like & 0.2 & 15 & Uncorrelated & $0.141^{+0.005}_{-0.005}$ & $75.070^{+0.183}_{-0.179}$ & $3.906^{+0.148}_{-0.144}$ \\
        Entire & Dipole-like & 0.2 & 20 & Uncorrelated & $0.148^{+0.005}_{-0.005}$ & $75.018^{+0.172}_{-0.175}$ & $4.081^{+0.146}_{-0.146}$ \\
        Quarter & Dipole-like & 0.2 & 12 & Uncorrelated & $0.188^{+0.005}_{-0.005}$ & $75.224^{+0.182}_{-0.180}$ & $5.469^{+0.178}_{-0.177}$ \\
        Quarter & Dipole-like & 0.2 & 15 & Uncorrelated & $0.191^{+0.005}_{-0.005}$ & $75.232^{+0.176}_{-0.180}$ & $5.575^{+0.181}_{-0.178}$ \\
        Quarter & Dipole-like & 0.2 & 20 & Uncorrelated & $0.195^{+0.004}_{-0.004}$ & $75.062^{+0.153}_{-0.158}$ & $5.749^{+0.163}_{-0.156}$ \\
        Entire & Disc-cone & 0.2 & 20 & Correlated & $0.195^{+0.005}_{-0.005}$ & $75.057^{+0.168}_{-0.175}$ & $5.768^{+0.185}_{-0.179}$ \\
        Entire & Disc-cone & 0.2 & 15 & Correlated & $0.196^{+0.005}_{-0.005}$ & $75.038^{+0.171}_{-0.172}$ & $5.793^{+0.188}_{-0.183}$ \\
        
        Entire & Disc-cone & 0.2 & 10 & Correlated & $0.198^{+0.005}_{-0.005}$ & $75.038^{+0.171}_{-0.175}$ & $5.880^{+0.189}_{-0.186}$ \\
        Entire & Dipole-like & 0.2 & 10 & Correlated & $0.197^{+0.005}_{-0.005}$ & $75.063^{+0.169}_{-0.169}$ & $5.849^{+0.175}_{-0.168}$ \\
        Entire & Ice-cream & 0.2 & 10 & Correlated & $0.196^{+0.005}_{-0.005}$ & $75.070^{+0.166}_{-0.171}$ & $5.812^{+0.173}_{-0.169}$ \\
        Entire & Ice-cream & 0.15 & 10 & Correlated & $0.146^{+0.005}_{-0.005}$ & $75.088^{+0.219}_{-0.224}$ & $5.753^{+0.226}_{-0.223}$ \\
        Entire & Ice-cream & 0.1 & 10 & Correlated & $0.097^{+0.005}_{-0.005}$ & $75.119^{+0.316}_{-0.325}$ & $5.644^{+0.348}_{-0.340}$ \\
        \hline
        \end{tabular}
    \end{threeparttable}
    \label{tab:Posterior distribution of signal parameters in different cases}
\end{table*}

\subsection{Biased extraction with Uncorrelated Modeling}
\label{sec:bias_from_uncorrelated}
We first present results from the uncorrelated modeling. We extract the 21\,cm signal from mock spectra measured with the badly designed disc-cone antenna beam for $N_r=20$, 15 and 12, respectively, 
using observation data taken at all positions in each orbit. 
We plot the recovered signals and the input signal in the left panel of Figure~\ref{fig:recovered_signal_nocov_comparison}, and list the fitted parameters in Table~\ref{tab:Posterior distribution of signal parameters in different cases}. Compared to the results using polynomial modeling in \citet{Shi2022}, we find that the fidelity of extraction is greatly improved using the modeling with $N_r=20$ and $15$, but the recovered signal is biased in a way that underestimates the signal strength. This implies that uncorrelated modeling might result in model bias. For $N_r<15$, e.g., the $N_r=12$ case corresponding to the purple regions in left panel of Figure \ref{fig:recovered_signal_nocov_comparison}, the extraction performance futher deteriorates owing to the insufficient characterization of the foreground sky.

Then we try to extract the 21\,cm signal from the mock spectra measured with the dipole-like antenna beam. Again we assume $N_r=20$, 15 and 12, and use observation data taken at all positions in each orbit. 
The recovered signal is shown in the middle panel of Figure~\ref{fig:recovered_signal_nocov_comparison}, and the fitted parameters are listed in Table~\ref{tab:Posterior distribution of signal parameters in different cases}. Even though the antenna is achromatic in this case, the recovered signal is still underestimated, showing that the under-estimating bias is not from the chromatic beam effect.

Similar biased fitting results have also been found in \citep{Anstey2023,Hibbard2023ApJ} when observations from multiple Local Sidereal Time (LST) bins are used. However, the fitting results remain unbiased with the same modeling \citep{Anstey2023,Hibbard2023ApJ} when observations from one single (or one averaged) LST bin are used. 

In the case of the DSL mission, to mimic observations with a single LST bin, one can use observation points within a quarter orbit, 
where the variations of response functions are relatively small. We repeat the above extraction process, and plot the recovered signal using observation points within a quarter orbit in the case of the dipole-like antenna beam in the right panel of Figure~\ref{fig:recovered_signal_nocov_comparison} and list the fitted parameters in Table~\ref{tab:Posterior distribution of signal parameters in different cases}. An unbiased 21\,cm signal is recovered in these cases. 
Unfortunately, using observation points within a quarter orbit, the unbiased fitting results for the 21\,cm signal is fake, because the foreground parameters are overfitted.

\begin{figure*}[ht]
    \centering
    \begin{subfigure}{0.45\textwidth}
        \centering
        \includegraphics[width=\linewidth]{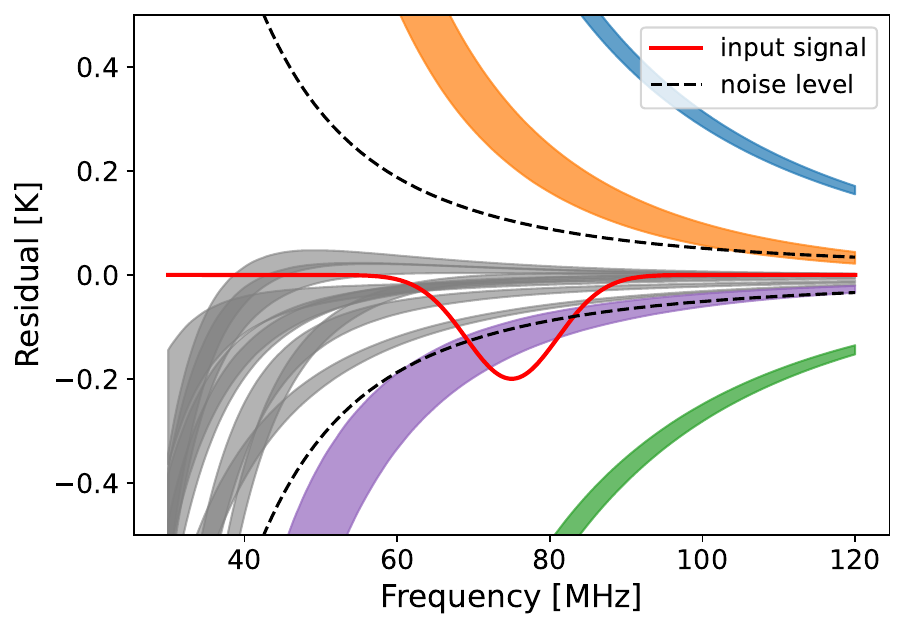}
    \end{subfigure}
    \begin{subfigure}{0.45\textwidth}
        \centering
        \includegraphics[width=\linewidth]{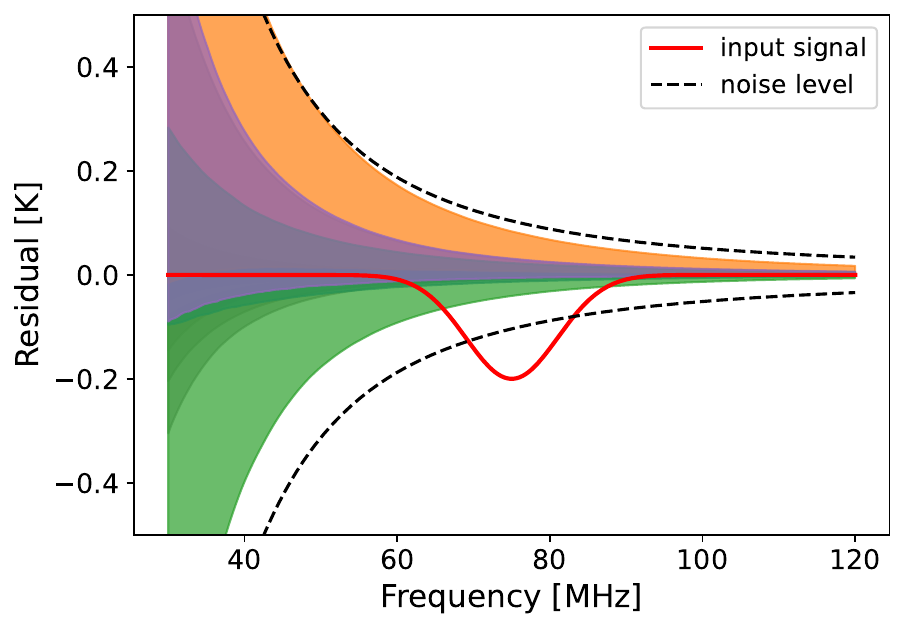}
    \end{subfigure}
    \caption{The residuals of foreground fitting at observation points inside and outside the quarter orbit, assuming the spectral index map is non-uniform (left) and uniform (right), respectively. Colored regions represent the 1$\sigma$ residuals at observation points outside the quarter orbit, while the grey regions represent the 1 $\sigma$ residuals at observation points inside the quarter orbit. The black dashed line is the thermal noise level in one observation, and the red line is the input 21\,cm signal with amplitude $A=0.2$ K.}
    \label{fig:demonstration of model bias and correlation2}
\end{figure*}

\subsection{Origin of the fake unbiasedness by observations within a quarter orbit}
\label{fake_unbiasedness}

In this section, we are going to show that when only the uncorrelated modeling of the residuals is applied, the unbiased fitting results using observation points within a quarter orbit are actually fake. 
The origin of the bias is the coupling between the time-varying response function and the non-uniformity of the sky.

We apply the fitted foreground parameters using observation points within a quarter orbit in Section~\ref{sec:bias_from_uncorrelated} to reconstruct the modeled spectrum data $T^{\rm 1/4\, rec}_{\rm model}(t,\nu)$ at several observation points within one full orbit. We compute 
the residuals between the modeled data using the reconstructed spectral indices $T^{\rm 1/4\, rec}_{\rm model}(t,\nu)$ and the {\it true} mock data $T_{\rm{mock}}(t,\nu)$. The 1$\sigma$ range in the fitted $\beta_p$ projects to the 1$\sigma$ residuals. 
The 1$\sigma$ residuals for observation points within the one quarter orbit are plotted with gray shades
in the left panel of Figure~\ref{fig:demonstration of model bias and correlation2}, and the residuals for four observation points outside the quarter orbit are shown with blue, orange, purple, and green areas, respectively. We find that the 1$\sigma$ residuals at observation points outside the quarter orbit are beyond the thermal noise level indicated by the dashed lines, 
while 1$\sigma$ residuals at observation points inside the quarter orbit are below the thermal noise level. This indicates that these foreground parameters fitted from the observation points within the quarter orbit are overfitted and unsuitable for all other observation points.

\begin{figure*}[htbp]
    \centering
    \includegraphics[width=0.96\textwidth]{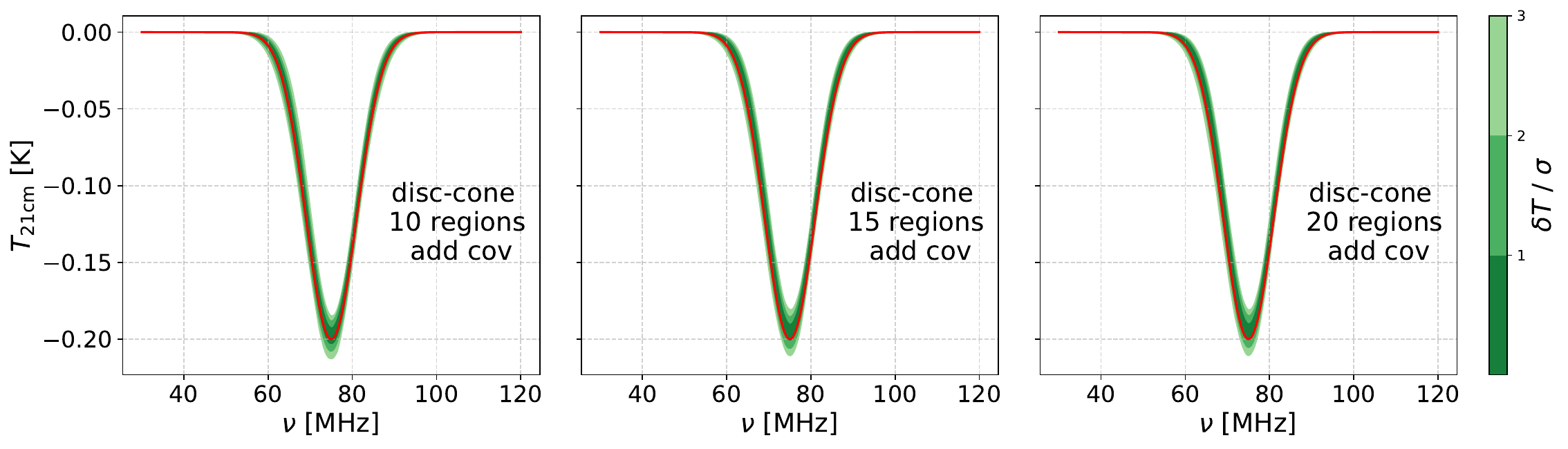}
    \caption{Same as Figure~\ref{fig:recovered_signal_nocov_comparison}, but using correlated modeling with $N_r=$10 (left), 15(middle), and 20 (right) to fit mock data convolved with a disc-cone antenna, respectively.}
    \label{fig:recovered_signal_addcov_10&20regions_200cmbeam_0o2K}
\end{figure*}

To further explain the origin of overfitting of foreground parameters, we simulate mock data with a sky map whose spectral index is uniform within each region and then use uncorrelated modeling to extract the 21\,cm signal. Similarly, we plot the residuals for data taken at different observation positions in the right panel of Figure~\ref{fig:demonstration of model bias and correlation2} and find that all the residuals are below the thermal noise level. This test confirms the origin of the overfitting as from the non-uniformity of the spectral index of the sky.

During the extraction process, the non-uniformity of spectral indices can be mitigated by increasing the number of celestial regions $N_r$. As anticipated, the extraction fidelity improves with larger $N_r$, as seen in Figure~\ref{fig:recovered_signal_nocov_comparison}, and the highest Bayesian evidence in our tests is found in $N_r=35$. However, a small bias still exists. In principle, the bias, or the overfitting, would only be  eliminated if the uncorrelated modeling could provide a complete description of the true spectral indices, similar as the case shown in the right panel of Figure \ref{fig:demonstration of model bias and correlation2}.

Therefore, a similar approach of using uncorrelated modeling is very likely to fail when applied to the real data of the DSL mission. Below, we show that by using the \textit{correlated modeling}, the underlying model bias can be avoided even for observations taken through the \textit{whole orbit}.

\begin{figure*}[htbp]
    \centering
    \includegraphics[width=0.96\textwidth]{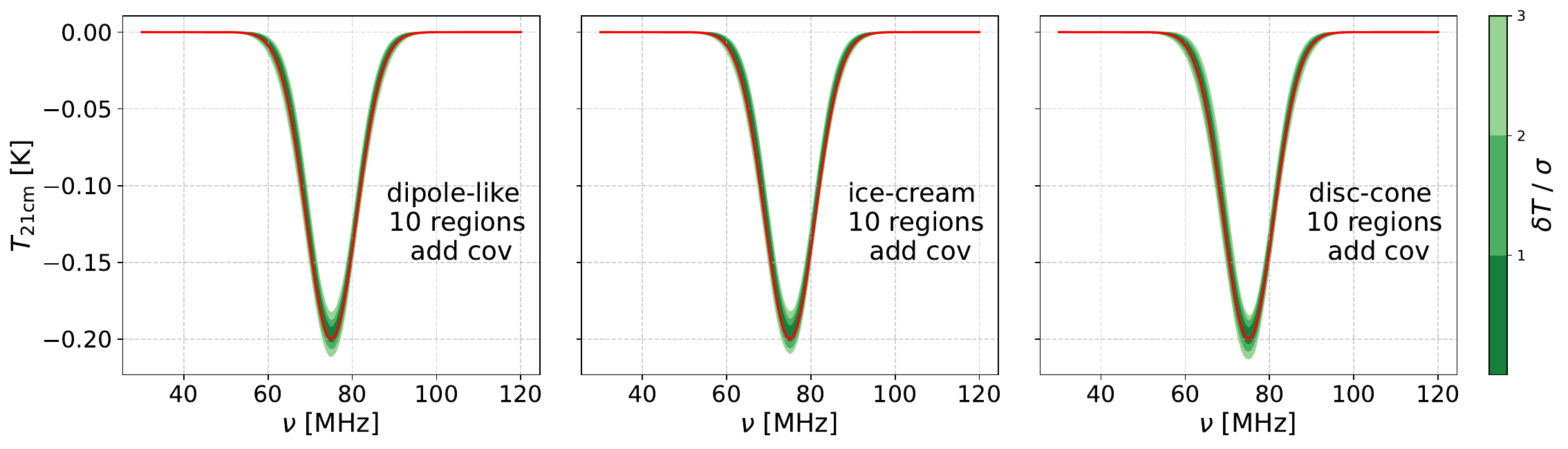}
    \caption{Same as Figure~\ref{fig:recovered_signal_nocov_comparison}, but using correlated modeling with $N_r=$10 to fit mock data convolved with a dipole-like antenna (left), an ice-cream antenna (middle), and a disc-cone antenna (right), respectively.}
    \label{fig:recovered_signal_addcov_10regions_0o2K_diff_beam}
\end{figure*}

\subsection{Unbiased extraction with correlated Modeling}
\label{sec:results_correlated_modeling}

In this section, we present results using correlated modeling for model errors.
The number of regions $N_r$ that the whole sky is divided into determines the complexity of the foreground model. In principle, a larger $N_r$ means a more complex and accurate model, but more computational cost is required. 
We first estimate the impact of sky division, i.e., the value of $N_r$, on the goodness of the signal extraction.

We vary the value of $N_r$ and extract the 21\,cm signals from the mock spectra with the disc-cone antenna beam. We plot the recovered signals with $N_r=$ 10, 15, and 20 in the left, middle, and right panels of Figure~\ref{fig:recovered_signal_addcov_10&20regions_200cmbeam_0o2K}, respectively. We also list the fitted parameters in Table~\ref{tab:Posterior distribution of signal parameters in different cases}. We find that using the correlated modeling, we can get unbiased extractions of the 21\,cm signal, which means our method is robust against model bias. We have tested all values of $N_r$ from 10 to 20, and do not find a significant downgrade in the goodness of extraction with $N_r=10$, compared to the extraction with $N_r=20$. The highest Bayesian evidence is reached with $N_r=18$. However, the final extraction results for the 21 cm signal with different $N_r$ are almost the same, similar to the results in Figure \ref{fig:recovered_signal_addcov_10&20regions_200cmbeam_0o2K}. Therefore, in the following, we perform signal extractions with $N_r=10$ to reduce computational costs. Note that when applied to real data, the true signal remains unknown, and we will need the Bayesian evidence to determine the most  appropriate $N_r$ in the modeling.

\begin{figure*}
    \centering
    \begin{subfigure}{0.32\textwidth}
        \centering
        \includegraphics[width=\linewidth]{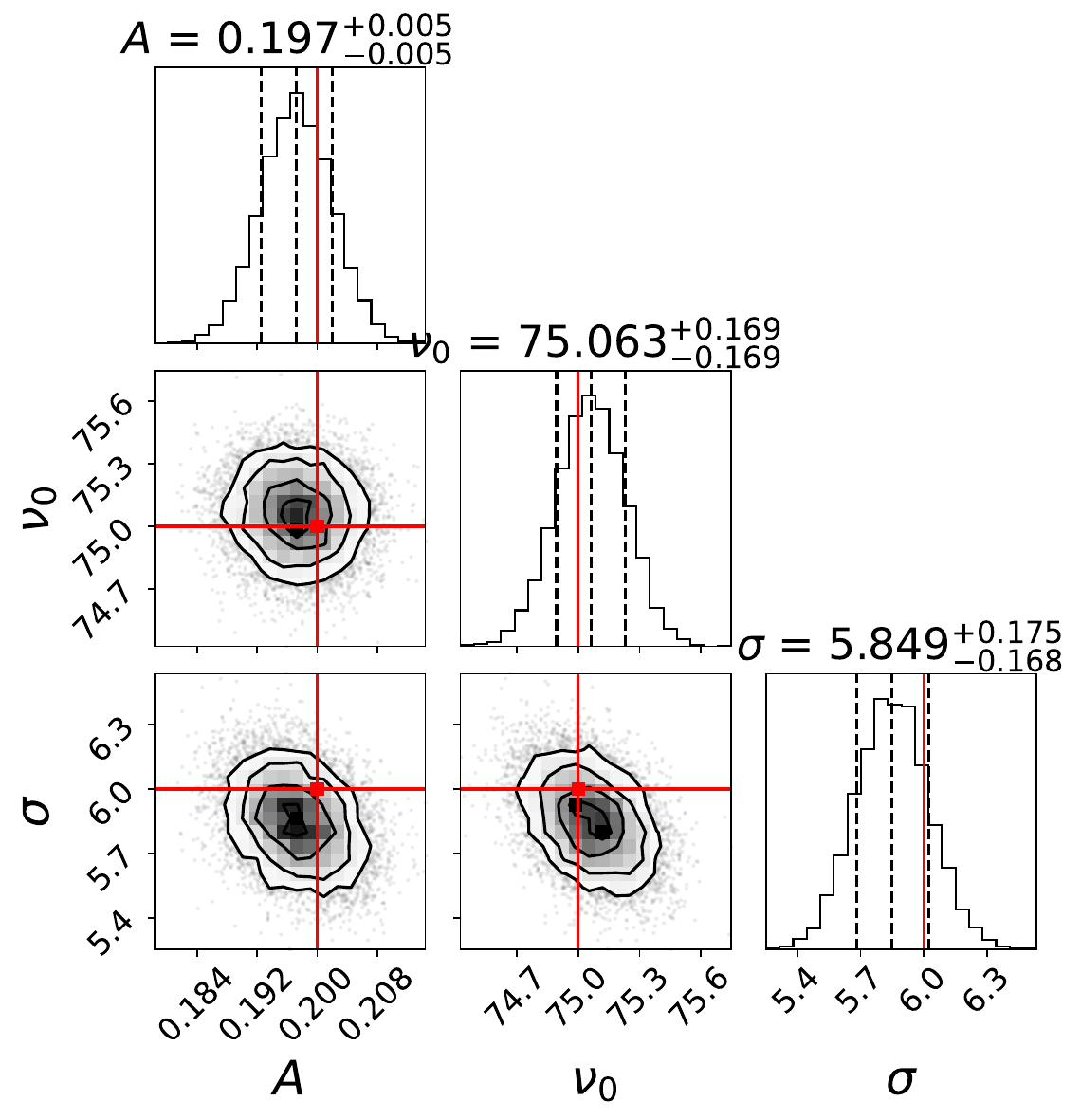}
    \end{subfigure}
    \begin{subfigure}{0.32\textwidth}
        \centering
        \includegraphics[width=\linewidth]{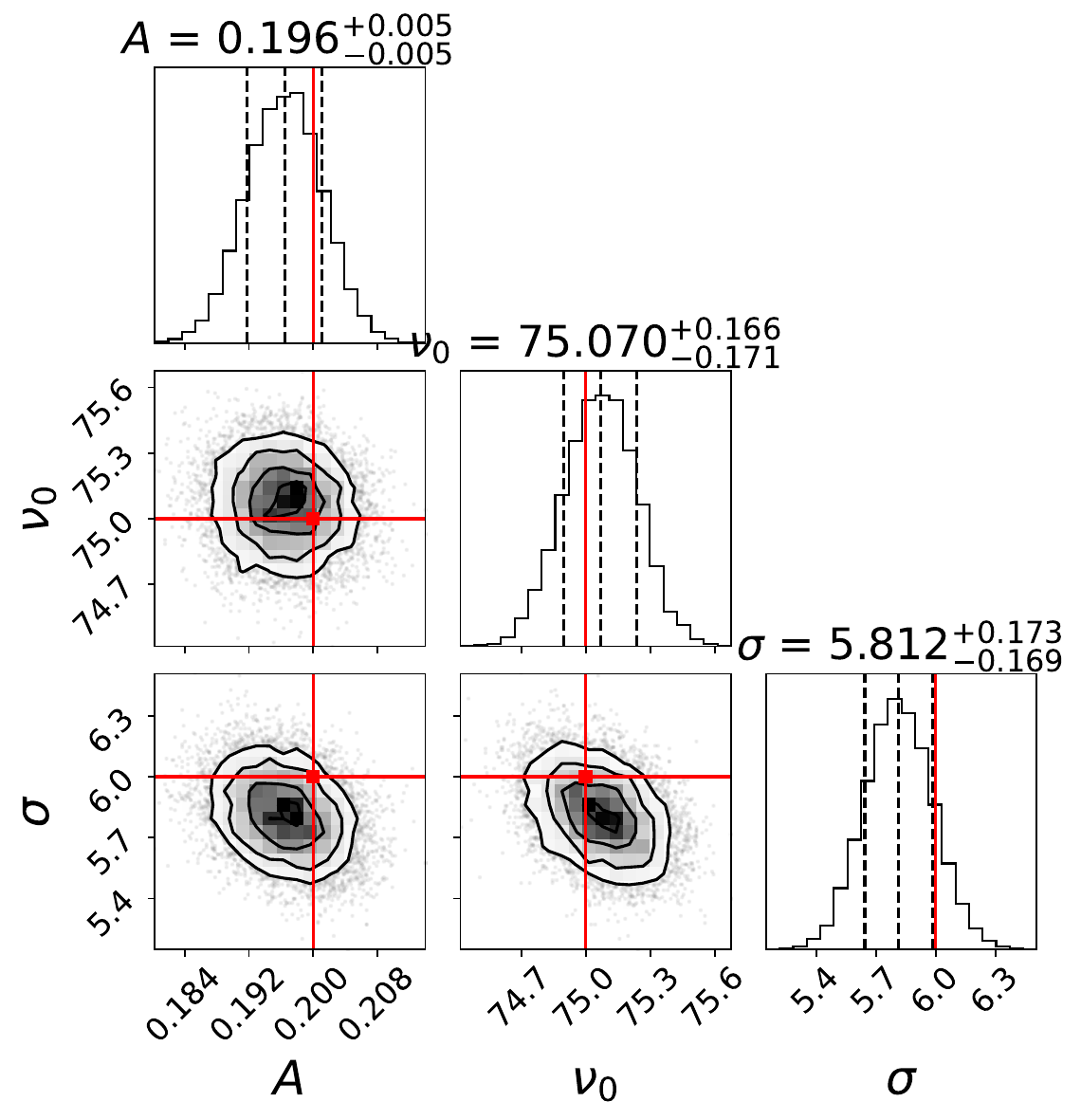}
    \end{subfigure}
    \begin{subfigure}{0.32\textwidth}
        \centering
        \includegraphics[width=\linewidth]{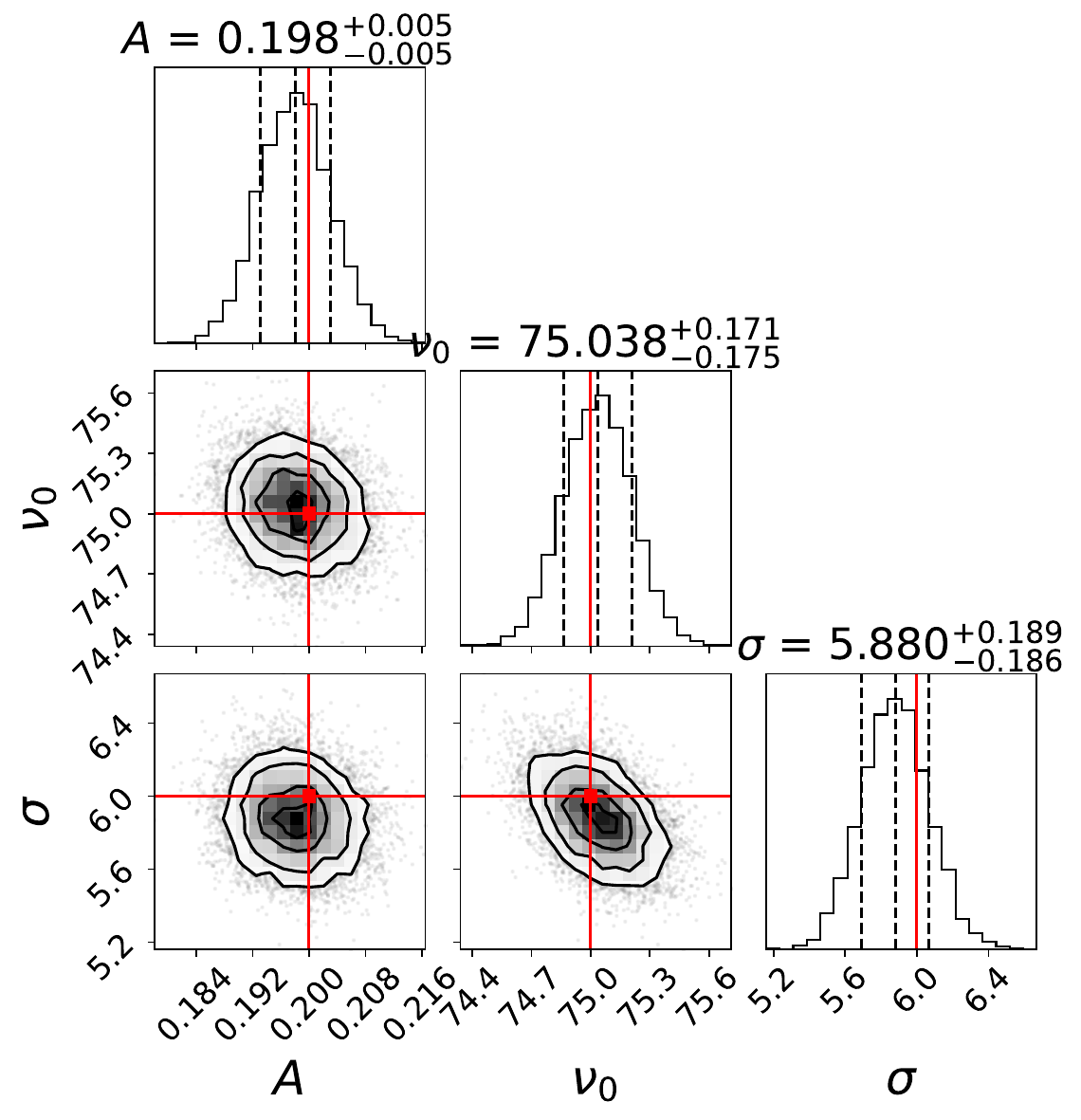}
    \end{subfigure}
    \caption{The posterior distribution of extracted signal parameters, corresponding to the cases in Figure~\ref{fig:recovered_signal_addcov_10regions_0o2K_diff_beam}. The red solid lines represent the input 21\,cm signal. The black dashed lines in each marginal distribution indicate the value of $16\%$, $50\%$, and $84\%$ quantiles, respectively.}
    \label{fig:corner_comparision}
\end{figure*}

Next, to test the robustness of our improved method against beam chromaticity, we explore how antenna beam models affect the goodness of extraction.
We vary the antenna beam models and extract the 21\,cm signals from the mock spectra with weaker chromatic antenna beams. 
The recovered signals extracted from the mock data convolved with the dipole-like antenna, the ice-cream antenna, and the badly designed disc-cone antenna are plotted in the left, middle, and right panels of Figure~\ref{fig:recovered_signal_addcov_10regions_0o2K_diff_beam}, respectively. We also list the fitted parameters in Table~\ref{tab:Posterior distribution of signal parameters in different cases}, and plot the posterior distributions in Figure~\ref{fig:corner_comparision}. In our example, the dipole-like antenna is perfectly achromatic, the disc-cone antenna is highly chromatic, while the chromaticity of the ice-cream antenna is intermediate between them. We do not find remarkable changes in the goodness of extraction among these three cases, which indicates that our method is robust against the chromatic beam effect.

Although the maximum level of the 21\,cm signal in standard cosmology is around $0.2$ K \citep{Xu2021ApJ}, there is still a large parameter space which would result in a weaker signal, which might be harder to extract. Therefore, here we also test shallower trough situations to check whether our method could still give good fitting results.

\begin{figure}
    \centering
    \includegraphics[width=0.96\columnwidth]{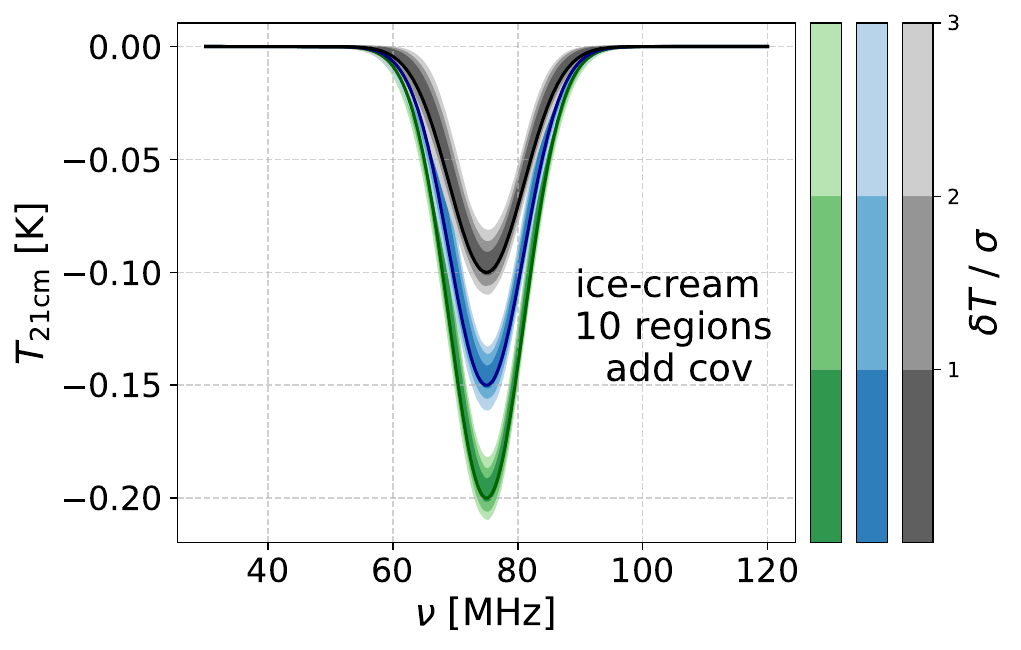}
    \caption{Same with Figure~\ref{fig:recovered_signal_nocov_comparison}, but using correlated modeling with $N_r=10$ to fit mock data convolved with an ice-cream antenna. We vary the amplitude $A$ of the input 21\,cm signal with 0.10 K (black), 0.15 K (blue), and 0.20 K (green), respectively.} 
    \label{fig:recovered_signal_addcov_10regions_icecreambeam_diff_amplitude}
\end{figure}

Here, we use the ice-cream antenna, which is a realistic design, and the antenna used in the DSL mission is likely to have similar performance.  We set $N_r=10$, and vary the amplitude of the signal, with $A=0.15$ K and $0.1$ K. The central frequency and the width of the mock signal remain unchanged, with $\nu_0=75$ MHz and width $\sigma_0=6$ MHz. We plot the recovered signals in Figure~\ref{fig:recovered_signal_addcov_10regions_icecreambeam_diff_amplitude}, and list the fitted parameters in Table~\ref{tab:Posterior distribution of signal parameters in different cases}. We find that the amplitudes $A$ are well-fitted with similar uncertainty levels for all cases. However, the uncertainty levels of the fitted central frequencies and the widths slightly increase with lower input amplitudes.

\section{Conclusion}
\label{sec:conclusion}

In this paper, we propose a physics-motivated error modeling framework to extract the 21\,cm global spectrum with the DSL mission. Our method builds on the previous sky division strategy in \citet{Anstey2021,Anstey2023} to account for spatial variations in the foreground spectral indices. We improve the uncorrelated Gaussian likelihood with a simulation-based covariance to include underlying correlations between residuals across observation points and frequencies.

We examine the uncorrelated modeling and the improved correlated modeling with mock data for the DSL mission. We find that with the correlated modeling, we successfully mitigate the model bias inherent in the uncorrelated modeling, which fails to extract unbiased signals from observations in one entire orbit with a limited number of foreground parameters. We demonstrate that this improved correlated modeling reliably recovers the input signal with a limited number of spectral index parameters ($N_r\gtrsim 10$), exhibits stability across different antenna types, and performs well in the extraction of weaker signals.

The current study adopts several simplifying assumptions. We have adopted a perfectly spherical Moon with negligible reflectivity.
It has been demonstrated that pure specular reflection by the Moon will have negligible impact on the 21 cm global spectrum extraction \citep{VZOP2024DSL}, but the Moon surface may have some level of diffuse reflection at the relevant frequencies.
Meanwhile, we have assumed rotationally symmetric beam patterns for all the antennas here. However, we have tested a case with a highly non-symmetrical beam, for which the response has a sharp decrease in a quarter of the beam pattern, and found that the 21 cm signal can still be reliably recovered as long as the beam pattern in known.
In addition to the fully-known beam shapes, we have also assumed a well-understood base pattern in the foreground temperature map though the spectral indices are set unknown and allowed to vary pixel-by-pixel, and the absence of point source contamination. These limitations will be discussed in future works. Overall, our framework provides a step forward towards a practical solution for unbiased 21\,cm global signal extraction, for both space-borne and ground-based experiments in the presence of beam chromaticity.

\section*{Acknowledgements}
We thank Haoran Zhang, Shenzhe Xu and Jiajun Yuan for providing the simulated beam profiles of the antennas.
This work was supported by National Key R\&D Program of China No. 2022YFF0504300, NSFC Grant No. 12361141814, China's Space Origins Exploration Program Nos. GJ11010401 and GJ11010405, the Chinese Academy of Sciences grant No. ZDKYYQ20200008, and the Specialized Research Fund for State Key Laboratory of Radio Astronomy and Technology.

\section*{}
{\Large{\it{Software:}}} {\tt{POLYCHORD}} \citep{2015MNRAS.450L..61H, 2015MNRAS.453.4384H}, healpy (\url{https://github.com/healpy/healpy}).


\FloatBarrier
\bibliography{21cm_cov}{}
\bibliographystyle{aasjournalv7}


\end{CJK*}
\end{document}